\documentclass[preprint]{aastex}
\usepackage{emulateapj5}
\usepackage{natbib}
\newcommand{\HI}{\ion{H}{1}}
\newcommand{\HII}{\ion{H}{2}}
\newcommand{\kms}{\mbox{km~s$^{-1}$}}
\newcommand{\Msol}{\mbox{M$_\odot$}}

\slugcomment{To appear in the Publications of the Astron.\ Soc.\ Aus.}
\shorttitle{Millimetre Science with ATCA}

\begin{document}

\title{Millimetre Science with the Upgraded Australia Telescope}
\author{T. Wong\altaffilmark{1} and A. Melatos\altaffilmark{2}}

\altaffiltext{1}{CSIRO Australia Telescope National Facility, PO Box
76, Epping NSW 1710, Australia; e-mail: Tony.Wong@csiro.au}
\altaffiltext{2}{School of Physics, University of Melbourne, Parkville
VIC 3010, Australia; e-mail: a.melatos@physics.unimelb.edu.au}

%
\begin{abstract}
A new astronomical window into the southern skies has been opened with
the high-frequency upgrade to the Australia Telescope Compact Array
(ATCA), which allows radio-interferometric mapping of sources at
wavelengths as short as 3mm.  In anticipation of the upgrade's
completion, a two-day workshop was held at the University of Melbourne
in November 2001.  The workshop covered a diverse range of fields,
tied together by a common theme of identifying key areas where ATCA
observations can have an impact.  More than half of the talks were
concerned with molecular clouds and star formation, with the remainder
covering topics such as molecular gas in the Galactic Centre, Seyfert
nuclei, and high-redshift objects.  Some early results from the 3mm
and 12mm prototype systems were also presented.  In consultation with
the speakers, we are presenting in this article a summary of the
talks.  The original slides are available from the ATNF website.

\end{abstract}

\keywords{instrumentation: interferometers, ISM: molecules, stars:
formation, galaxies: ISM, galaxies: high-redshift, masers}

%
%
\section{Introduction}

Millimetre (mm) astronomy has grown rapidly in the past decade, spurred by
advances in receiver sensitivity and growing scientific interest in
what might be termed the ``cold universe.''  While molecular lines are
still the ``bread and butter'' of millimetre astronomy, a wide range
of continuum sources can also be studied at these wavelengths, and
today mm astronomy spans nearly the full range of astrophysical
research, from solar-system studies to cosmology.

The success of centimetre-wave interferometry, demonstrated by
facilities such as the Westerbork Synthesis Radio Telescope (WSRT),
led to the development of millimetre interferometers in the 1970's and
1980's.  There are currently four dedicated mm arrays, all located in
the Northern Hemisphere at latitudes of 35\arcdeg\ to 45\arcdeg.  A
much larger array, to be built in Chile by an international
consortium, will provide a manyfold increase in collecting area over
all existing mm arrays put together.  This Atacama Large Millimeter
Array (ALMA) will be completed around 2011.  In the meantime, however,
another millimetre array is making its debut in the Southern
Hemisphere: the Australia Telescope Compact Array (ATCA), being
upgraded to operate at wavelengths as short as 3mm, is poised to open
a new astronomical window into the southern skies.

In anticipation of the completion of the ATCA upgrade, a workshop on
millimetre-wave science was convened at the University of Melbourne on
29-30 November 2001.  Participants from the Australian astronomical
community, as well as guests from overseas, came to discuss and learn
about areas of science that could be addressed with the new ATCA.  As
is inevitable with a small meeting, many areas could not be properly
represented---including, but not limited to, cosmology and VLBI---so
this summary should be considered only a sampling of what might be
possible.  We also apologise in advance if, in trying to summarise the
meeting in a cohesive way, we have been rather selective in our
recollections.  Slides from the talks are available for download from
the ATNF
website\footnote{http://www.atnf.csiro.au/whats\_on/workshops/mm\_science2001}.

\section{The ATCA as a Millimetre Interferometer}

The Australia Telescope Compact Array, situated near Narrabri, New
South Wales (longitude 150\arcdeg\ E, latitude 30\arcdeg\ S) consists
of six 22m diameter Cassegrain antennas.
Five of the antennas are transportable along a 3km long east-west
track---now supplemented by a 200m north spur extending from the
centre of the track---and are being outfitted with new 3mm (85-105
GHz) and 12mm (16-26 GHz) receiver systems.  The sixth antenna is
fixed at a station 3 km west of the track's end and will likely be
outfitted with a 12mm receiver only.  At the time of the meeting, a
3-element prototype system was available at frequencies of 85-91 GHz
and 16-22 GHz; for the current status of the upgrade see the 
online documentation for 
observers\footnote{http://www.atnf.csiro.au/observers/}.

Designed primarily as a cm array, the ATCA suffers from several problems
as a mm array: a small primary beam, few baselines, poor aperture
efficiency, and a far from ideal site at an elevation of 200m.  In
addition, the present correlator provides a maximum bandwidth of 128
MHz at each of two frequencies, much less than the 1 GHz or more that
is typical at mm wavelengths (see Table~\ref{tbl:arrays}).  
The velocity range that can be achieved by
placing the two frequency windows side-by-side (allowing
for some overlap) is limited to $\sim$600
\kms\ at 90 GHz.  This limitation will be overcome once a new wideband
(2 GHz) FX correlator is installed, as described during the meeting by
{\bf W. Wilson}.  That project, to be undertaken in conjunction with
development for the Square Kilometre Array (SKA), is 
due for completion around 2006.

\begin{table*}
\begin{center}
\caption{Upgraded ATCA Compared with Other Millimetre Arrays\label{tbl:arrays}}
\bigskip
\begin{tabular}{lcccrccccc} \hline
 Array & Altitude & $n$ & $D$ & $nD^2$ & $\eta_A$
& No. of & Freq.~Range & BW & $b_{\rm max}$ \\ 
 & (km) & & (m) & (m$^2$) & & Polns & (GHz) & (GHz) & (km) \\ \hline
 ATCA & 0.2 & 5 & 22 & 2420 & 0.4 & 2 & 16--26, 85--105 & 0.3 & 3.0 \\ 
 BIMA & 1.0 & 10 & 6.1 & 370 & 0.7 & 1 & 70--116, 210--270 & 1.6 & 1.5 \\
 IRAM & 2.6 & 6 & 15 & 1350 & 0.7 & 1 & 80--116, 210--250 & 1.0 & 0.4 \\
 NMA & 1.3 & 6 & 10 & 600 & 0.6 & 1 & 85--116, 213--237$^a$ & 1.0 & 0.4 \\
 OVRO & 1.2 & 6 & 10.4 & 650 & 0.7 & 1 & 86--116, 210--270 & 4.0 & 0.4 \\\hline
\end{tabular}
\end{center}
$^a$Also 126--152 GHz.\\
KEY: $n$---number of antennas;
$D$---diameter of each antenna;
$\eta_A$---aperture efficiency of each antenna;
BW---continuum bandwidth after combining both sidebands or IFs, if available;
$b_{\rm max}$---maximum baseline.
\end{table*}

Despite the ATCA's limitations as a millimetre array, once the initial
stage of the upgrade (fitting complete receiver systems on 5 antennas)
is complete in 2003, it will possess several unique strengths:

\begin{itemize}

\item Enormous collecting area, giving it an effective aperture
($\eta_A n D^2$) comparable to that of the Plateau 
de Bure interferometer.

\item Ability to observe two polarisations simultaneously.

\item Wide-bandwidth mixers based on indium phosphide MMIC technology,
which need only be cooled to about 20 K, as opposed to 4 K for 
typical SIS junctions. 

\item A maximum baseline of 3 km, allowing resolution as fine as 
0\farcs2 once radiometric phase correction is implemented.

\end{itemize}

As discussed at the meeting by {\bf T. Wong}, mm interferometry is much more
challenging than cm interferometry, as a result of more stringent
technical requirements (e.g. surface accuracy, pointing, and baseline
determination) and the strongly variable effects of the atmosphere.
In particular, fluctuations in the H$_2$O vapour layer lead to
fluctuations in atmospheric opacity and observed interferometric phase
that must be monitored with frequent calibrations.  These fluctuations
tend to be more severe in the daytime and in summer, so winter nights
are expected to offer the best observing conditions.  As the
fluctuations occur on a range of size scales, longer baselines will
suffer from greater atmospheric phase noise under the same observing
conditions.  Long-baseline experiments also suffer from a lack of
absolute flux calibrators, since the planets, which are the best
primary flux calibrators, are likely to be resolved out.

The effects of atmospheric phase noise can be largely undone if the
relative H$_2$O vapour column above each antenna in the direction of
the source is known.  One way to determine the water vapour column is
to accurately measure the strength of an H$_2$O emission line, such as
that at 22.2 GHz.  This ``water vapour radiometry'' technique was
discussed by {\bf R. Sault} and is currently being tested at the ATCA,
with some encouraging results.  The system observes the 22 GHz line in
four channels to better isolate line from continuum emission and
reduce systematic errors.

\begin{figure*}
\begin{center}
\includegraphics[height=5.5in,angle=-90]{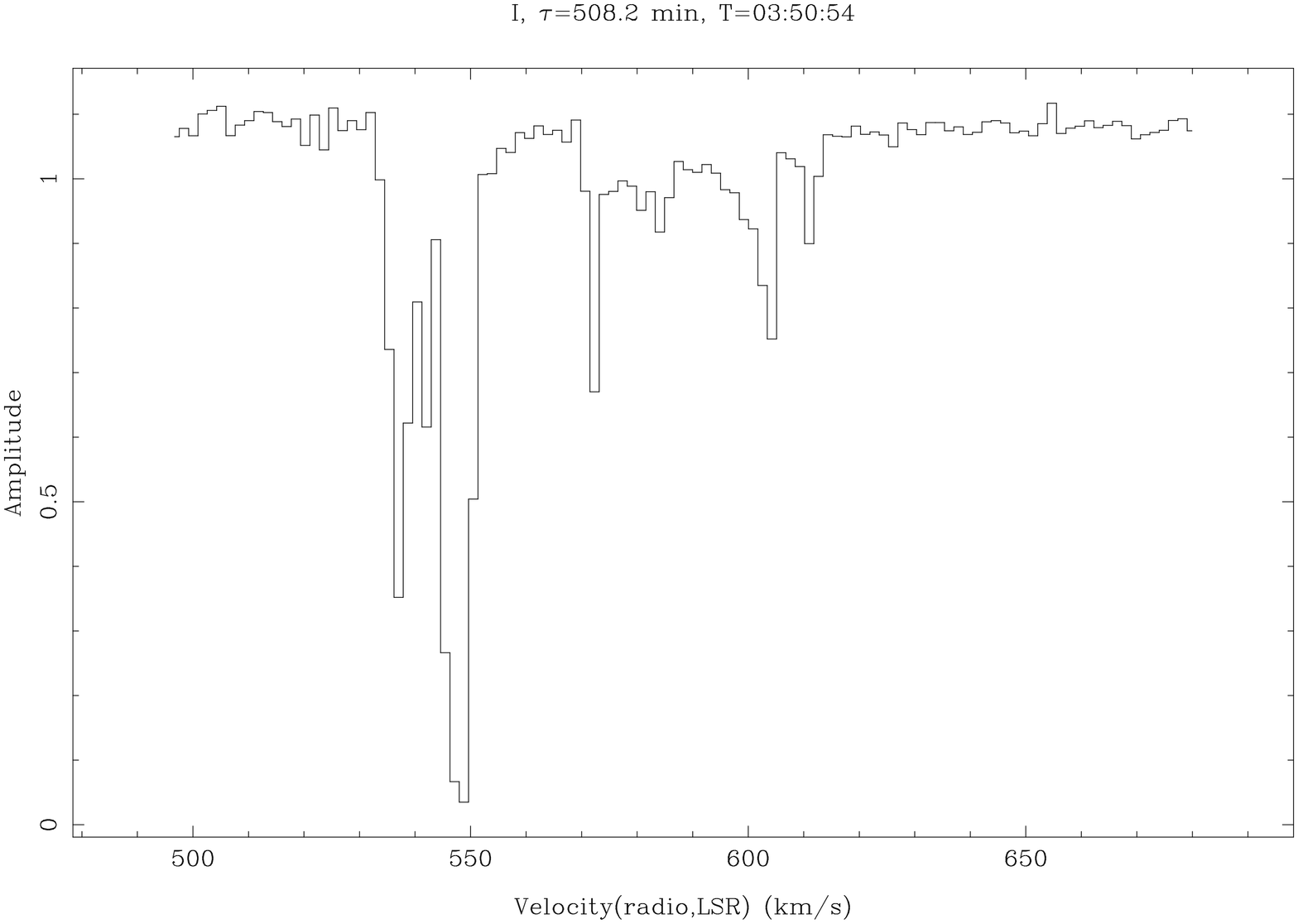}
\caption{
Absorption spectrum of the HCO$^+$ (0--1) line in front of the
continuum source in Centaurus A.  Three hours of on-source integration
were obtained with baselines of 46, 77, and 122m on 6 October 2001.
All three baselines have been averaged together to produce the final
spectrum.  The amplitude scale is in arbitrary units.
\label{fig:cena}}
\end{center}
\end{figure*}

{\bf B. Koribalski} presented some of the first science observations
made with the prototype 3mm system, beginning with ``first light'' in
2000 November.  These included observations of molecular clouds in the
Milky Way and LMC and detection of dense molecular gas in nearby
galaxies like NGC 253 (Figure~\ref{fig:n253}) and Circinus.  The
highlight thus far has been the absorption spectrum of Centaurus A in
the HCO$^+$ and HNC lines, which provides the first $\sim$1\arcsec\
resolution measurement of gas motions in front of the core of this
radio galaxy (Figure~\ref{fig:cena}).  A search for CO absorption in
the quasar PKS 1921-293 was also conducted.

\vskip 0.21truein
\includegraphics[width=3.25in]{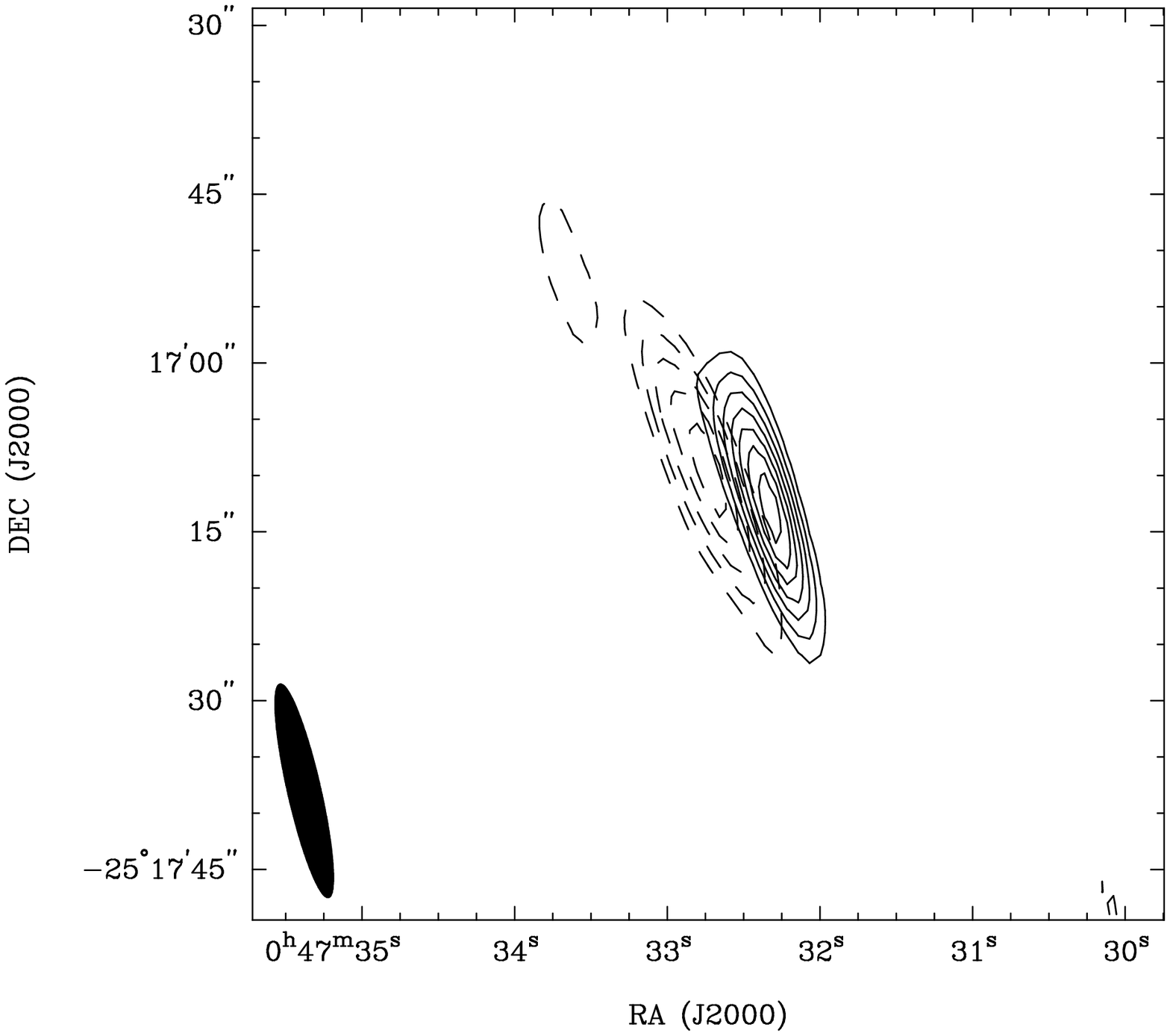}
\figcaption{
Deconvolved image of the nucleus of NGC 253 in the HCO$^+$ (1--0) line.
The beam size in 19\farcs4 $\times$ 2\farcs9.  Redshifted emision
($V_{\rm LSR} > 230$ \kms) is shown as solid contours, and blueshifted
emission ($V_{\rm LSR} < 230$ \kms) as dashed contours.
The data were taken with the 3-element system on 5 October 2001 over
a period of 5.5 hours, with about two hours of on-source integration.
\label{fig:n253}}
\vskip 0.21truein

{\bf R. Manchester} presented the first image of the SN 1987A remnant at
17--19 GHz, made with the 3-element 12mm system in 2001 October.  The
image, based on a 12-hour synthesis, shows a $\sim$20 mJy source that
is moderately resolved with a synthesised beam of 2\farcs8 $\times$
1\farcs8 (see Figure~\ref{fig:sn1987a}).  A fit of a thin shell to the
$uv$ data gives a diameter of 1\farcs5 and a fit of a Gaussian gives a
FWHM of 1\farcs1.  An IAU Circular \citep{Manchester:01} has been 
released, and follow-up observations at higher resolution are planned.

\vskip 0.25truein
\includegraphics[height=3.25in,angle=-90]{sn1987a.eps}
\figcaption{
Image of the remnant of SN 1987A derived from a 10-hour synthesis at
the ATCA.  Data from two frequencies (17 and 19 GHz) have been combined.
The synthesised beam (2\farcs8 $\times$ 1\farcs8) is shown at the
lower right.  The contour levels are $-1$, 1, 2, 5, and 10 mJy 
beam$^{-1}$.  Courtesy R. Manchester. 
\label{fig:sn1987a}}
\vskip 0.25truein

\section{Molecular Clouds and Star Formation}

Studies of molecular clouds and star formation within them occupy a
prominent place in millimetre astronomy, and were the subject of much
discussion during the workshop.  Indeed, as one of the ATCA's greatest
attractions lies in its southern location, studies of Galactic star
formation are poised to benefit immensely---most of the Galaxy's
nearby star-forming regions (e.g. Chamaeleon, Corona Australis, Lupus,
Ophiuchus), not to mention the Galactic Centre, are located in the
southern sky.

While almost all of the gas in molecular clouds consists of H$_2$ and
He, these species do not emit radiation under conditions typically
found in molecular clouds.  Except in rare cases where H$_2$
absorption against a UV source can be measured, the amount of
molecular gas must be inferred indirectly.  A common technique is to
observe emission lines from tracer molecules like CO that have millimetre
or submillimetre rotational transitions.  Not only can these lines
provide information on the distribution and velocity of the molecular
gas, but different molecular tracers can be used to probe different
density regimes within a cloud.

The ATCA's ability to study {\it nearby} molecular clouds is limited by its
small field of view, lack of sensitivity to extended structure, and
inability to observe the bright CO(1--0) line at 115 GHz (this last
constraint may be overcome in a future upgrade).  Thus, studies of the
fragmentation of molecular clouds into clumps and cores
\citep[cf.][]{Williams:00} are best undertaken with other instruments.
Better suited to the ATCA will be studies of more distant clouds, such as
in the Galactic Centre and Magellanic Clouds, and studies of dense
molecular cores where star formation is occurring or likely to occur.

\subsection{Molecular clouds near the Galactic Centre}

{\bf Cornelia C. Lang} proposed several experiments that the ATCA can
perform to increase our understanding of the Galactic Centre.
Long-term monitoring of the millimetre flux from the radio source Sgr
A$^\ast$, thought to be associated with the central black hole, will
help to determine whether the circumnuclear accretion disk is
transitory or stable.  Existing NH$_3$, HCN, and HCO$^+$ images of the
circumnuclear disk, taken with the VLA and BIMA interferometers, show
filamentary streamers of molecular gas falling onto the disk,
providing direct evidence of inward gas transport
\citep{McGary:01,Wright:01}.  However, they are inconclusive on the
issue of disk stability.

Single-dish mapping of molecular gas in the vicinity of the Galactic
Centre Radio Arc has revealed that it is physically related to the
ionised gas in this region as well as to the radio non-thermal filaments
(NTFs).  Recently, interferometric studies made with OVRO reveal a
complex geometrical arrangement and kinematic structure in the
molecular gas (Lang, Goss \& Morris 2002, submitted).  The associations
between the ionised and molecular gas suggest that the action of
energetic winds from young stellar clusters (such as the Arches
cluster) plays an important role in this region.  The NTFs are thought
to trace a large-scale, poloidal field normal to the Galactic plane in
the inner 250 pc of the Galaxy.  The interfaces between the NTFs and
molecular clouds may be sites where twisting and reversal of magnetic
field lines occurs, followed by magnetic reconnection which
accelerates the electrons to relativistic speeds
\citep[e.g.,][]{Serabyn:94}.  With its superior view of the Galactic
Centre, high spectral resolution, and dual polarisation capability,
the ATCA can help elucidate these processes.

Millimetre images of the Galactic Centre can be compared with the
recent Chandra Galactic Center Survey \citep{Wang:02}, which covers the
central 300 $\times$ 100 pc with a resolution ($\sim$1\arcsec) that is
well matched to the ATCA.\@  Such multiwavelength studies will help to
resolve puzzles like the origin of the diffuse Fe line emission at 6.4
keV from the G.C. molecular clouds: is this fluorescent line excited
by an energetic outburst from Sgr A$^\ast$ \citep{Koyama:96}, or from
a more local source of X-rays responsible for driving the expanding
molecular shells in the region \citep{Martin:99}?

\subsection{Shocked molecular gas\label{sec:shock}}

The structure and chemical composition of a shock in the ISM are
controlled by the shock speed $v_s$.  In J-type, or `fast', shocks,
which occur for $v_s \gtrsim 40\,{\rm km\,s^{-1}}$, molecular gas is
dissociated and ionised to a substantial degree immediately upstream
and downstream from the shock, both by collisions and by UV radiation
produced at the shock front.  It is only far downstream from the
shock, where the gas cools to $\approx 500\,{\rm K}$, that molecule
reformation occurs.  J-type shocks are inefficient emitters of
infrared line radiation, but strong emitters of far-infrared continuum
(because the UV flux is reprocessed by grains).  On the other hand, in
C-type, or `slow', shocks, which occur for $v_s \lesssim 40\,{\rm
km\,s^{-1}}$, the shock front is broadened by ambipolar diffusion, and
molecules are not dissociated to the same extent.  The energy radiated
by such shocks appears mainly in infrared lines (atomic and ionic
fine-structure lines and molecular rotation-vibration transitions).
Since the molecules are not destroyed, they can reach much higher
temperatures than in J-type shocks \citep{Kaufman:96}, driving
endothermic reactions and hence interesting chemistry.

An important consequence of C-shocks discussed by {\bf M. Wardle} at
the meeting is the efficient production of H$_2$O from atomic oxygen
(O $\rightarrow$ OH $\rightarrow$ H$_2$O).  This can give rise to
intense water maser emission at 22 GHz and at submillimetre
wavelengths \citep{Kaufman:96}.  Within supernova remnants (SNRs), a
strong X-ray flux releases high-energy electrons which can excite the
Lyman and Werner bands of H$_2$.  The resulting far-UV flux can
dissociate some of the H$_2$O back into OH \citep{Wardle:99}, thereby
accounting for the strong OH maser activity at 1720 MHz associated
with SNRs \citep{Frail:94}.  The associations between shocked H$_2$
and OH masers in the SNRs G359.1-0.5 \citep{Lazendic:02} and
G349.7+0.2 (Lazendic et al., in preparation) are consistent with this
picture, but direct observations of H$_2$O, using the ATCA at 22 GHz,
would be valuable.

Wardle also suggested a number of lines in the 3mm band which could be
used to probe shock regions.  Good targets are the interaction regions
between a supernova remnant and a cloud; many are distant, so a
30\arcsec\ field of view is acceptable.  If X-ray induced dissociation
of water is indeed occurring, one would also expect the SO$_2$/SO
ratio to be affected, since SO$_2$ is more easily dissociated than SO.  This
could be tested by observations of the SO($3_2$--$2_1$) transition at
99.3 GHz and the SO$_2$($3_{13}$--$2_{02}$) transition at 104 GHz.
With the existing prototype system, the ATCA can already image the
shock-tracing SiO(2--1) line at 86 GHz.  The presence of SiO in the
gas phase is an indication of grain sputtering \citep{Schilke:97}.

\subsection{Identifying molecular cores}

As reviewed by {\bf T. Bourke}, molecular cores have typical masses of
1--10 \Msol\ and sizes of 0.1--0.5 pc.  They therefore occupy a small
fraction of the mass and an even smaller fraction of the volume of an
entire cloud.  Since cores with embedded young stellar objects (YSO's)
will be heated, they can be detected in wide-field maps of molecular
clouds in infrared or submillimetre dust emission.  Such maps will be
provided by focal plane arrays on single-dish telescopes, such as the
SEST Imaging Bolometer Array (SIMBA), and by an upcoming Legacy
project on the Space Infrared Telescope Facility (SIRTF).  With its
high resolution and sensitivity, the ATCA will be important for
detailed studies of the cores identified by such surveys.

{\it Starless} cores, which by definition are not (yet) infrared
sources, are probably best identified using line emission from dense
gas tracers such as NH$_3$ (23 GHz), CS (98 GHz), and N$_2$H$^+$ (93
GHz).  The system of NH$_3$ lines is also useful for determining
densities and temperatures within cores.  Temperature diagnostics are
essential because observed linewidths may be dominated by turbulent
motions.

While nearby clouds like Taurus are important laboratories for
studying low-mass star formation, identifying sites where {\it
massive} ($>$10 \Msol) star formation is occurring is considerably
more difficult, due to the rapid evolutionary timescales for massive
stars, their tendency to form in clusters where source confusion is a
problem, and their relative scarcity (hence, they tend to be observed
at larger distances).  The basic technique, as outlined by Bourke, is
to search for cores with dense gas (as traced by CS emission) and hot
dust (as traced by IRAS far-infrared fluxes) where the massive
protostar has not yet ionised its surroundings (i.e., no detectable 5
GHz emission from an \HII\ region).  A useful starting point would be
the catalogue by \citet{Bronfman:86}.

Masers are another important signpost for massive star formation, as
pointed out by {\bf V. Minier}.  Class II methanol masers, especially the
lines at 6.7 and 12.2 GHz, appear to be strongly correlated with the
{\it early} stages of massive star formation, namely outflows and hot
molecular cores, but maser activity apparently peaks before the
emergence of an ultracompact \HII\ region \citep{Minier:01}.

\subsection{From dense cores to protostellar disks}

The formation of protostellar disks is still poorly understood.
Schematically, it appears that dense cores form within molecular
clouds (perhaps at the intersections of filamentary structure) and
then collapse gravitationally to form rotationally supported disks.
Yet the dynamical processes and timescales involved in disk formation
are still quite uncertain.  A useful idealised model (neglecting
rotation and magnetic fields) is the ``inside-out'' collapse model of
\citet{Shu:77}, which represents the collapse of a singular isothermal
sphere.  This model predicts power-law density and velocity profiles
of $n(r) \propto r^{-1.5}$ and $v(r) \propto r^{-0.5}$ in the inner,
collapsing regions \citep[see][]{Evans:99}.

{\bf M. Hunt} described how millimetre spectroscopy can be used to determine
chemical abundances and physical conditions in molecular cloud cores.
She presented the results of single-dish studies of southern molecular
clouds with Mopra and SEST.\@  Of particular interest are molecules that
can be observed at several optically thin transitions, so that a
``rotation diagram'' \citep{Goldsmith:99} can be used to infer the
kinetic temperature, under the assumption that the level populations
are in local thermodynamic equilibrium (LTE).  Corrections for
departures from LTE can be made using the large velocity gradient
(LVG) approximation.  Among the potential tracers for ATCA studies are
CH$_3$OH (at least 7 thermal transitions from 86--115 GHz), HC$_3$N (3
transitions), and OCS (3 transitions).  The ATCA will be able to perform 
these studies at much higher resolution, reducing the uncertainties
associated with averaging over a finite beam.

{\bf T. Bourke} reviewed the rapidly growing literature concerned with
spectroscopic infall signatures in dense cores.  The most common
signature, observed in optically thick lines like HCO$^+$, is an
asymmetric line profile skewed towards the blue, since the red side of
the line is formed at larger (hence cooler) radii from the central
source than the blue side \citep[see explanation in][]{Evans:99}.  In
addition, the presence of a cold, static envelope outside the collapse
region leads to a self-absorption feature near the systemic velocity.
Examples of line profiles that match these predictions are given in
studies by \citet{Gregersen:97} of embedded YSO's and \citet{Lee:99}
of starless cores.  These authors also show that blue-skewed
profiles are not seen in optically thin tracers like N$_2$H$^+$ or
H$^{13}$CO$^+$, in agreement with infall models.

While most of these observations were made with single-dish
telescopes, high-resolution data are important in that they allow
better discrimination between the different effects (infall, rotation,
outflow, and turbulence) which can be superposed in a single-dish
beam.  Indeed, the best-known candidate for a collapsing protostar, in
the isolated globule B335 \citep{Zhou:93,Choi:95} looks less clear-cut
when viewed at higher resolution \citep{Wilner:00}.  This raises the
question of whether single-dish data alone can sufficiently constrain
the range of possible models.  To date, the most convincing evidence
for infall motions---line profiles with the ``inverse P Cygni'' shape
(blueshifted emission with redshifted absorption)---has been gathered
with interferometers \citep[e.g.,][]{Difrancesco:01}, where the beam
is sufficiently small to see the infalling gas in absorption.

{\bf G. Blake} pointed out that
high angular resolution is also needed to study the transition from
the embedded phase, where most of the gas is in an infalling envelope,
to the T Tauri phase, where most of the gas is in a rotating disk, as
the angular size scales involved are $\sim$1000 AU, or 7\arcsec\ at
the distance of Taurus.  This transition appears to occur quite
rapidly, given the apparent paucity of transitional forms
\citep[one example is proposed by][]{Hogerheijde:01}.  Disk envelopes
can be traced via the HCO$^+$ and HCN lines, both of which are
accessible with the ATCA prototype system, while the disk itself
should be traceable in continuum emission, especially once the
wideband correlator is available.

\subsection{Modelling disk evolution}

As reviewed by {\bf S. Maddison} at this meeting, simulations are an
important tool for understanding the evolution of protostellar disks.
Massive disks are susceptible to gravitational instabilities, which
can induce mass accretion and hence feed back on the gravitational
potential.  These effects are difficult to model analytically but are
well-suited for N-body simulations.  An important observational input
is the mass surface density of the disk, which can be crudely
estimated from the mass and radius.  In the early stages of accretion,
when the disk is massive, gravitational instabilities dominate and can
lead to spiral arm formation or disk fragmentation---which can in turn
lead to binary or giant planet formation.  In later stages, most of
the mass has accreted to the centre, the disk rotation is nearly
Keplerian, and accretion is driven by turbulent viscosity.

A nice example of concordance between theory and observation occurs in
the case of circumbinary disks.  Subarcsecond-resolution
observations of GG Tau \citep{Guilloteau:99} show a central hole in
the dust distribution, consistent with the gap that should be cleared
by resonant torques in a binary system.  Observations at still higher
resolution, however, will be needed to test the more detailed
predictions of the simulations, such as the formation of bisymmetric
spiral structure and accretion streams feeding the stars.

\subsection{Disks around pre-main-sequence stars\label{sec:pms}}

{\bf G. Blake} reviewed observations of protoplanetary disks, such as
those associated with T Tauri stars.  With radii of order 100 AU
(1\arcsec\ at the distance of Taurus), subarcsecond resolution will be
needed to study such disks.  Such resolution has recently been
achieved by other mm arrays \citep[e.g.,][using the IRAM Plateau de
Bure]{Simon:01}, and should be achievable by the ATCA once the radiometric
phase correction system is operational.  Even with a small number of
baselines and relatively incomplete coverage of the visibility plane,
the basic structure of disks can be investigated by modelling the
visibilities directly \citep[e.g.,][]{Guilloteau:98}.

Molecular lines can be a valuable probe of disk kinematics, but the
large ranges in density and temperature around YSO accretion disks
lead to a complex chemistry which is vital to understand.  An
important phenomenon is the depletion of molecules onto dust grains
that occurs primarily at high densities and low temperatures but can
be reversed by irradiation and shocks.  The commonly employed
molecular tracer CO, for instance, may be depleted even when H$_2$
remains in the gas phase.  Such ``freeze-out'' is especially likely
near the disk midplane, where densities are highest.  Where CO is
detectable, it is likely to be optically thick and trace only the
disk's outer layers; observing rarer CO isotopes, such as $^{13}$CO
(110.2 GHz) and C$^{18}$O (109.8 GHz), will be crucial for probing the
vertical structure of disks.  This will be a key focus for future
high-resolution studies, and a strong argument for 110 GHz capability
at the ATCA.

Another way to trace the disk kinematics is via maser emission.  The
relative fringe phase between maser spots is insensitive to baseline
errors and atmospheric effects \citep{Wright:83}, and can be used to
deduce their relative positions to hundredths of an arcsecond if the
spots are separated in velocity \citep[e.g.,][]{Plambeck:90}.  In
addition, once the instrumental baseline has been determined by
observing quasars across a wide range of hour angles and declinations,
absolute position information is also available
\citep[e.g.,][]{Forster:78}.

For deducing physical quantities such as density, temperature, and
chemical composition, observations of multiple lines, including (for
density and temperature) transitions between different energy levels
of the same molecule, are necessary.  Because of the ATCA's limited
frequency range, many of the commonly used molecular tracers will only
be observable in a single transition (generally the $J = 1 \rightarrow
0$ line).  Nonetheless, an interferometer map may still provide useful
constraints on source geometry to complement single-dish measurements.
Examples of studies that have exploited this method include
\citet{Dutrey:97} and \citet{Hogerheijde:97}.  Infrared spectra can
also provide complementary information on physical conditions, with
the advantage that many lines can be observed simultaneously in a
given spectrum \citep[see e.g.\ the review by][]{vanDishoeck:98}.

\subsection{Bipolar Outflows}

Outflows from star-forming cores are ubiquitous, as pointed out by
{\bf T. Bourke}, but are often so embedded in dust that they can only be
studied in the radio.  Fine velocity resolution is critical, as the
outflow speeds can be as little as a few \kms.  Spatial resolution is
essential as well, especially in crowded regions where identifying the
source of the outflow can be problematic.  Some of the obvious
southern targets are regions around $\rho$ Ophiuchi, R Coronae
Australis, and Circinus.

While CO is the usual tracer because of its high abundance and low
critical density, other molecular lines (e.g. from CS or NH$_3$) can
provide additional information.  Of particular importance for the ATCA in
the near term is the thermal SiO line at 86.8 GHz, which falls
within one of the frequency windows available to the 3-element
prototype system.  The SiO molecules are liberated from dust grains
shocked by the outflow, and thus trace the outflow more directly than
a general gas tracer like CO.  The presence of an outflow is often
indicated in single-dish spectra by broad line wings at relative
velocities of up to $\sim$60 \kms.

Although the origin of bipolar outflows remains controversial
\citep[see reviews by][]{Shu:00,Konigl:00}, there is little doubt that
magnetic fields play an important role in accelerating and collimating
the observed jets.  Magnetic fields in dense molecular regions are
usually revealed by the linear polarisation of thermal emission from
spinning dust grains.  In regions where the continuum is weak, it is
still possible to detect polarisation in molecular line transitions
due to the Goldreich-Kylafis \citep{Goldreich:81} effect, although
this effect is rarely observed.  For recent detections see
\citet{Greaves:99} and \citet{Girart:99}.  With its exceptional
sensitivity and dual polarisation capability, the ATCA should be in a
good position to exploit this technique.

{\bf R. G. Smith} stressed the importance of complementary infrared
observations.  IR absorption spectra are sensitive to the presence of
ices, since the same molecule produces different absorption profiles
in the gas vs.\ solid phase.  Optical and near-infrared polarimetry
can be used to reveal light that has been scattered by a disk.  These
techniques were illustrated in the case of OH231.8+4.2, an evolved
star in which numerous molecular lines have been observed
\citep{Morris:87,Sanchez:00}.  Of particular interest among these is
HCO$^+$, which appears to be enhanced in the lobes of the outflow,
perhaps due to shock-induced reactions.

\subsection{Debris Disks}

Late in their evolution, protostellar disks have dispersed most of
their gas and a {\it debris disk} remains, with a mass only a few
percent of the central star.  Observations of such disks (and their
immediate precursors, the disks around T Tauri and Herbig Ae/Be stars)
were reviewed by {\bf Chris Wright}.  Debris disks were first identified as
excess far-infrared emission around main sequence stars like Vega
($\alpha$ Lyrae) and $\beta$ Pictoris.  Ground-based (and more
recently HST) imaging and spectroscopy revealed dusty disks around
these stars, sometimes with warps and gaps indicative of planetary
perturbers.

A primary motivation for studying these systems is to determine the
timescale for disk dispersal, which is important for constraining
theories of planet formation.  This is best done by comparing the dust
emission from disks in various stages of evolution
\citep[e.g.,][]{Holland:98}.  Although inferring dust masses from
millimetre or submillimetre continuum emission is hardly
straightforward, one can get a handle on the grain properties using
mid-infrared spectroscopy of the 10$\mu$m and 20$\mu$m silicate
features.

Studying the disk rotation and gas-phase chemistry requires molecular
line observations.  However, searches at millimetre wavelengths for
gas in debris disks have generally been unsuccessful
\citep[e.g.,][]{Liseau:98}.  Direct observations of H$_2$ with the
Infrared Space Observatory (ISO) indicate that there is indeed
molecular gas present \citep{Thi:01}, so CO must be strongly depleted
(by 2 to 3 orders of magnitude) compared to typical regions of
molecular clouds.  The reason for this depletion could be a
combination of photodissociation and freeze-out onto grains.  In
addition, at temperatures above 20 K, the $J=2$ and higher levels of
CO are preferentially populated compared to $J=1$, and the CO(1--0)
line is thus relatively weak \citep{Liseau:98}.  Indeed, the disk
around the nearby T Tauri star TW Hya has been detected more strongly
in the higher $J$ transitions \citep{Kastner:97}, with no CO(1--0)
detected at Mopra (upper limit $\sim$0.3 K in a 30\arcsec\ beam).
These results suggest that heating of the disk is rather efficient,
probably due to flaring of the outer parts of the disk \citep[cf.\ HST
observations by][]{Burrows:96}.

\section{Interstellar and Circumstellar Masers\label{sec:masers}}

\subsection{Methanol masers\label{sec:methanol}}

Methanol (CH$_3$OH) masers are generally found in active star-forming
regions, and have traditionally been divided into two empirical
classes.  Class I masers are well-separated from compact continuum
sources and OH masers.  Class II sources are generally associated with
ultracompact \HII\ (UCHII) regions, although this association has
become less convincing with higher resolution data.  Nonetheless,
since certain maser lines are only seen in Class I sources and the
rest only in Class II sources, the distinction between the two classes
appears to have a physical basis.  Methanol maser lines have been
detected at numerous frequencies from 6.7 to 157 GHz, with the
brightest lines being the Class II transitions at 6.7 and 12.2 GHz.
The maser ``spots'' themselves have sizes of order 1 milliarcsecond,
although they are typically found in clusters of up to an arcsecond in
size.

{\bf V. Minier} reviewed observations of the spatial distribution of
masers, many of them conducted at milliarcsecond resolution with
VLBI.\@ Interestingly, it appears that the 6.7 and 12.2 GHz masers are
often offset from their nearest UCHII regions by several arcseconds
\citep{Walsh:98,Minier:01}.  In these cases, they appear to coincide
instead with hot molecular cores (HMCs) seen in NH$_3$ and CH$_3$CN.
A possible evolutionary scenario was presented, in which methanol
masers form in the protostellar (HMC) phase, the young star begins to
ionise its surroundings and destroy the masers, and finally an UCHII
region emerges without maser activity.  Testing this scenario requires
resolving individual protostars using high-resolution instruments,
inferring their masses from continuum fluxes, and inferring their ages
by comparing observed spectral line ratios with chemical models
\citep[e.g.,][]{Rodgers:01}.  Much of this work should be possible
with the ATCA.

{\bf A. Sobolev} discussed how physical conditions can be inferred from
observations of molecular masers and thermal lines.  Besides CH$_3$OH,
a number of other maser transitions can be studied, including H$_2$O,
OH, SiO, and NH$_3$.  Among the transitions of particular interest for
the ATCA upgrade is the 95.2 GHz Class I methanol maser.  A southern
survey of these masers was recently conducted with the Mopra telescope
by \citet{Valtts:00}.  The spectra of methanol masers often show
multiple, independently varying components that are generally
attributed to motions within the molecular cloud, and can thus be
used to probe the turbulent velocity spectrum.  For example, the
positions and spectra of masers near 25 GHz detected in the Orion
Molecular Cloud \citep{Johnston:92} can be modeled in terms of
Kolmogorov-like turbulence, although the spectrum appears to fall more
steeply with wavenumber than the standard Kolmogorov law
\citep{Sobolev:98}.  Meanwhile, thermal lines, such as CS(2--1) at
96.4 GHz, SiO(2--1) at 86.8 GHz, and CH$_3$OH (2$_K$--1$_K$) at
96.7 GHz, provide useful information about temperature, density, and
shock chemistry, while high-resolution (interferometric) mapping can be
used to study jets and shocks.

A more detailed look at the excitation of Class II methanol masers was
provided by {\bf D. Cragg}.  Although the formation of inverted
populations (``pumping'') that can lead to masing remains an area of
ongoing investigation, Class II transitions are thought to be pumped
radiatively by infrared photons from the warm dust surrounding young
massive stars.  While the IR photons produce the population inversion,
bright masers are the result of high methanol abundance, together with
beaming and/or amplification of the UCHII background continuum.
\citet{Sobolev:94} and \citet{Sobolev:97a} have modeled these
processes in detail to reproduce the high observed brightness
temperatures of the 6.7 and 12.2 GHz masers.  Strong masing is found
to occur for dust temperatures $T_d > 150$ K, gas densities $10^6 <
n_{\rm H}\, [\rm cm^{-3}] < 10^8$, and methanol abundances relative to
H$_2$ of $> 10^{-7}$.  The models also predict Class II masers at
81.0, 85.6, 86.6, 86.9, 94.5, 108.9, and 111.2 GHz
\citep{Sobolev:97b}, in addition to the relatively strong line at
107.0 GHz, which has already been detected in over 25 sources
\citep{Valtts:99,Caswell:00}.  Observational progress in detecting new
maser transitions has been rapid, with recent detections of the 86.6
and 86.9 GHz masers in the prototypical Class II source W3(OH)
reported by \citet{Sutton:01} using BIMA, and of the 85.6, 86.6, and
86.9 GHz masers in the rich southern source G345.01+1.79 reported by
\citet{Cragg:01} using SEST and Mopra.

\subsection{Late-type variable stars\label{sec:miras}}

Molecular masers are also observed around late-type variable stars
such as Miras, which are characterised by pulsation periods in the
range of $\sim$100 to $\sim$500 days, strong molecular bands in their
optical spectra (TiO and VO for type M, ZrO for type S, CN and C$_2$
for type C), and bright Balmer emission lines.  The Balmer lines are
produced by strong atmospheric shocks driven by the stellar
pulsations; the outgoing, spherical pressure wave steepens to form a
shock as it propagates into a decreasing density gradient.  Mass loss
at a rate of up to $10^{-5}$ \Msol\ yr$^{-1}$ creates a
circumstellar envelope which hosts masers such as OH, H$_2$O, SiO, and
HCN.  The SiO and HCN masers are believed to arise close to the
stellar photosphere ($R \sim 10^{14}$ cm), whereas the H$_2$O masers
are associated with the expanding shell at distances $R \sim
10^{14}$--$10^{16}$ cm, and OH masers with the outer parts of the
shell at distances $R \gtrsim 10^{16}$ cm.

The light curves of late-type variables vary with wavelength, as
discussed by {\bf G. Rudnitskij}.  For example, the near-infrared
maxima lag behind the optical maxima by about 0.1$P$, where $P$ is the
period \citep{Lockwood:71}.  The H$_2$O maser flux also appears to lag
behind the optical flux, by about 0.05--0.35$P$ in RS Virginis
\citep{Lekht:01}.  Such a large phase lag appears to rule out
radiative pumping; instead, it is proposed that the arrival of a
periodic shock into the masing region leads to enhanced collisional
pumping of the maser \citep{Rudnitskij:90}.  Consistent with this
hypothesis is the presence of a peak in the cross-correlation function
at a lag of 3--5$P$ (where $P \approx 1$ yr), roughly the time for a
shock to travel from the stellar surface to the region where masers
can form ($R \approx 10^{14}$ cm).  Shock pumping is also suggested
for the H$_2$O maser flare in R Leo that was observed $\sim$18 months
after an H$\alpha$ flare \citep{Esipov:99}.  Although the H$_2$O and
SiO maser spots are expected to be unresolved by the ATCA, it was noted
that if they are sufficiently bright and appear at distinct
velocities, then their relative positions can be determined to a
fraction of a beam (\citealt{Plambeck:90}; see \S\ref{sec:pms}).

Rudnitskij closed his presentation with the provocative suggestion
that H$\alpha$ flares may be due to a Jovian-sized planet in an
eccentric orbit at a few AU, which has survived the red giant phase
and excites strong atmospheric shocks during each periastron passage.
If the stellar mass loss is asymmetric, a highly eccentric planetary
orbit can be maintained against the circularising action of viscous
torques.  

\section{Interstellar Medium in Galaxies\label{sec:exgal}}

{\bf S. Curran} noted that the ATCA is well suited to mapping the spatial
distribution of molecular gas within Seyfert galaxies.
Notably, the two nearest starburst/Seyfert hybrids, NGC 4945 and
Circinus, are only observable with southern telescopes.  An important
observation would be to demonstrate the ability of a dusty circumnuclear
torus to collimate a molecular outflow: \citet{Curran:99} argued for
the existence of such an outflow in Circinus based on SEST mapping,
but higher resolution observations are needed to confirm its
association with the observed ionisation cone \citep[][and references
therein]{Veilleux:97}.

Another unresolved issue centres on the question of how well the dense
gas tracer HCN correlates with star formation.  NGC 4945 shows a
higher HCN/CO ratio, and a higher HCN/FIR (far-infrared) ratio, than
Circinus despite lower gas densities ($\sim 10^4$ vs.\ $10^5$
cm$^{-3}$) \citep{Curran:01}.  In light of BIMA maps of the Seyfert
galaxy NGC 1068 which show that most of the HCN is concentrated
towards the centre, whereas CO and star formation are located in a
ring $\sim$1 kpc in radius \citep{Helfer:95}, this suggests that much
of the HCN emission in NGC 4945 comes from more diffuse gas associated
with the dusty torus in the inner $\sim$10 pc, whereas in Circinus it
is associated with star-forming cores in the $\sim$500-pc-radius
molecular ring.  However, better angular resolution will be needed to
test this hypothesis.  

{\bf M. Dopita} discussed the implications of recent theoretical work
on the predicted spectrum of dust emission at millimetre wavelengths.
First of all, while the dust absorption coefficient $\kappa$ is usually
assumed to scale as $\kappa \propto \nu^{0.7}$ at low frequencies,
implying a steeply declining emission spectrum with wavelength,
laboratory measurements by \citet{Mennella:98} indicate that
$\kappa$ (and hence the emissivity) is strongly temperature-dependent,
so that hot dust would show excess emission at long (millimetre)
wavelengths.  This could be an important consideration for
interpreting mm/sub-mm continuum fluxes from galactic nuclei.
Secondly, electric dipole radiation from small ($\sim$100 atoms)
aspherical, rotating grains is predicted to peak at a rest frequency
near $30\,{\rm GHz}$ \citep{Draine:98}.  This prediction follows from
a consideration of several effects including ``plasma drag'' (the
interaction of the grain's dipole moment with passing ions) and
``electrostatic focussing'' (the change in the collisional cross
section due to electrostatic attraction or repulsion), and may explain
the excess background emission near this frequency observed by
\citet{Leitch:97} and others.  Finally, magnetic dipole radiation can
be emitted by dust containing ferromagnetic materials
\citep{Draine:99}, with emission peaking in the tens of GHz (depending
on the characteristic gyrofrequency).

\section{High-Redshift Molecular Gas\label{sec:hiz}}

{\bf T. Wiklind} reviewed the observability of molecular gas at high
redshift.  Studies of emission and absorption are complementary, with
emission lines providing global properties (gas content and rotational
velocity) whereas absorption lines probe the gas on much smaller
scales and with much greater sensitivity.  Although the number of
detections to date is small, the fact that the CO line has already
been observed out to $z \sim 5$ has allayed concerns that low
metallicities would render it undetectable.  On the other hand, there
is little doubt that metallicity {\it does} affect the $L_{\rm
CO}/M_{\rm H_2}$ ratio \citep[e.g.,][]{Maloney:88}, so the objects
detected (especially in emission) are likely to be unusually luminous
and dusty.  Moreover, optimistic early assessments of the
detectability of CO at $z > 5$ \citep{Silk:97} have been questioned by
more recent simulations \citep{Combes:99}.

\begin{figure*}[b]
\begin{center}
\includegraphics[width=4.5in]{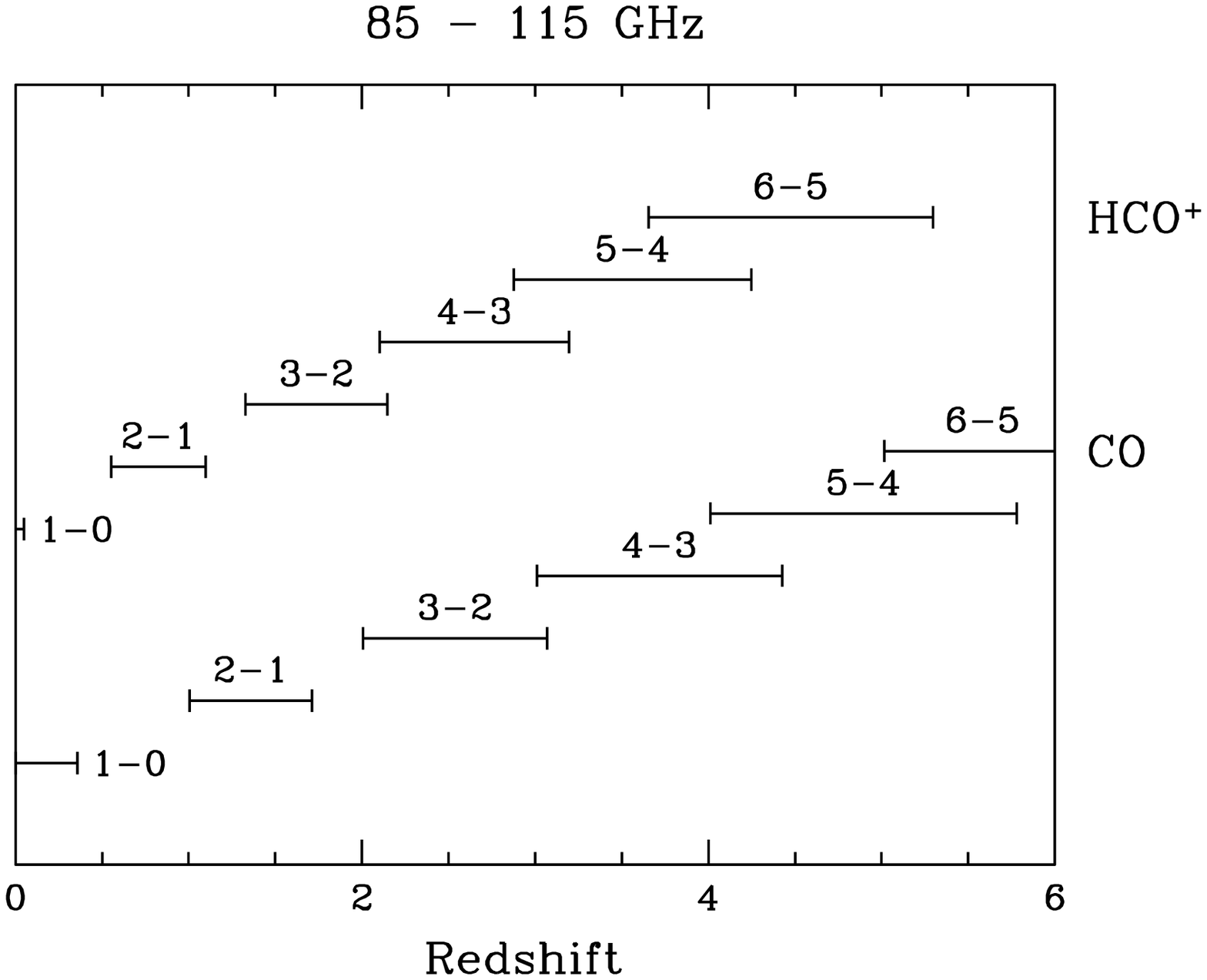}
\caption{
Observability of the CO and HCO$^+$ lines at high redshift with a
frequency coverage from 85--115 GHz.  For each transition there is 
a redshift range (given by a horizontal bar) where the line falls
within the observable band.  Adapted from a figure by T. Wiklind.
\label{fig:hiz}}
\end{center}
\end{figure*}

As of 2001 November, thirteen CO emission sources had been detected at
$z>1$ (all with interferometers, which provide the necessary gain
stability for long integrations); more than half of the detected
sources appear to be gravitationally lensed.  The highest redshift
source is at $z$=4.69 \citep{Omont:96}.  Note that because of the
increased cosmic microwave background (CMB) temperature at high redshift,
higher $J$ transitions of CO can be as bright or even brighter than
the $J$=1$\rightarrow$0 transition.  Thus, high-redshift sources can
still be detected in the 3mm band, as high $J$ transitions are
redshifted into this band.  Figure~\ref{fig:hiz} shows the redshift
coverage of the ATCA system, assuming its frequency range can be
extended to 115 GHz.

Absorption lines can be detected out to any redshift as long as a
suitably bright continuum source is available.  For instance, one can
achieve comparable signal-to-noise on the $z > 0.89$ quasar PKS 1830-211
\citep{Wiklind:96} as on a nearby galaxy like Cen A.  The integrated
opacity can be approximated as
\[\int \tau_\nu\,d\nu \propto \frac{N_{\rm tot}}{T_x^2} \mu_0^2\]
where $N_{\rm tot}$ is the total column density of the molecule, $T_x$
is the excitation temperature given by the level populations, and
$\mu_0$ is the permanent dipole moment.  This expression shows that
absorption is particularly sensitive to cold gas, and that a
molecule with a high dipole moment like HCO$^+$ can be as strong as CO
in absorption, despite lower abundance.  Including the HCO$^+$
series of lines allows for even more complete redshift coverage in the
3mm band (Figure~\ref{fig:hiz}).

With its excellent sensitivity, the ATCA should be a superb tool for
detecting molecules in high-redshift galaxies.  Potential emission
targets are best identified at submillimetre wavelengths
\citep[e.g.,][]{Frayer:98}, where the strongly increasing dust
emission with frequency offsets the dimming due to increased
cosmological distance \citep{Hughes:96}.  The SCUBA and SIMBA
instruments on the James Clerk Maxwell Telecope (JCMT) and the SEST
respectively are being utilised for such searches.  Redshifts can be
determined by several techniques, listed here in order of decreasing
accuracy: (1) optical spectroscopy \citep[e.g.,][]{Ivison:98}, (2)
multi-colour optical photometry \citep[e.g.,][]{Mobasher:96}, or (3)
the radio-to-submillimetre spectral index \citep{Carilli:99}.  While
optical redshifts (methods 1 and 2) are more accurate,
\citet{Smail:00} have concluded that most of the SCUBA-detected
sources are at $z \gtrsim 2$ and do not have optical identifications.
The third method, based on the assumption that the observed correlation
between radio and far-infrared emission in star-forming
galaxies \citep{Condon:92} is valid at high redshift, requires only flux measurements at 1.4
and 350 GHz.

Potential absorption targets can be selected from catalogues of
radio-loud (especially millimetre-loud) QSO's.  In principle, optical
and/or near-infrared colors can be used to select quasars which are
reddened by dust \citep{Webster:95}, as would be expected if there
were intervening molecular gas.  On the other hand, as noted by {\bf
R. Webster} and {\bf M. Drinkwater} at the meeting, synchrotron
emission may account for the reddened spectra in a large fraction of
cases \citep{Francis:01}, highlighting the need for more refined
selection criteria.  Yet another problem is choosing the observing
frequency in the absence of a reliable optical redshift (for either
the background source or the absorber).  Lacking redshift information,
it should still be possible to scan across the 3mm band in a
reasonable amount of time \citep[e.g.,][]{Wiklind:96}, particularly
once the ATCA's bandwidth has been upgraded.

An increased sample of molecular gas detections at high redshift would
benefit a variety of studies.  By providing accurate redshifts for
submillimetre sources, they would significantly improve our knowledge
of the evolution of massive galaxies and of the star formation rate.
In addition, by observing multiple transitions of the same molecule
and deriving the excitation temperature, one can place upper limits on
the CMB temperature as a function of redshift.

As discussed by {\bf M. Murphy} and {\bf M. Drinkwater} in the final
talk of the workshop, high signal-to-noise CO and \HI\ spectra of
absorption-line systems can be used to provide constraints on
variations in the fine-structure constant ($\alpha = e^2/\hbar c$)
with redshift.  Specifically, the ratio of the \HI\ hyperfine to the
CO rotational line frequency is proportional to $y = \alpha^2 g_p$,
where $g_p$ is the proton $g$-factor \citep{Murphy:01b}.  A recent
analysis by \citet{Murphy:01b} leads to an upper limit of $\Delta
\alpha/\alpha < 3 \times 10^{-6}$ at redshifts of 0.25 and 0.68,
assuming a constant $g_p$.  However, higher redshift systems are
needed to test claims, based on optical QSO spectra, that $\Delta
\alpha/\alpha$ deviates significantly from 0 at $z>1$
\citep{Webb:99,Murphy:01a}.

\section*{Acknowledgments}


We would like to thank the staff and students of Melbourne University for
helping to make the workshop a success.  We also appreciate the
generosity of the MNRF International Collaboration Committee, which
helped to support the travel costs of many of our visitors.

\end{document}